\documentclass[oneside]{elsart}
\usepackage{times}
\usepackage[T1]{fontenc}
\usepackage[latin1]{inputenc}
\usepackage{graphicx}

\makeatletter

\newcommand{\noun}[1]{\textsc{#1}}

\makeatother
\begin{document}
\journal{Nuclear Instruments and Methods B}

\begin{frontmatter} 

\title{An Algorithm for Computing Screened Coulomb Scattering in \noun{Geant4}\thanksref{g4}} 

\author{Marcus H. Mendenhall\corauthref{cor}\thanksref{mfel}}

\corauth[cor]{Corresponding author}

\thanks[g4]{This work not affiliated with or endorsed by the \textsc{Geant4} collaboration}

\thanks[mfel]{Supported by grant FA9550-04-1-0045 from the DOD MFEL Program}

\address{Vanderbilt University Free Electron Laser Center, P. O. Box 351816 Station B, Nashville, TN 37235-1816, USA} \ead{marcus.h.mendenhall@vanderbilt.edu}

\author{Robert A. Weller}

\address{Department of Electrical Engineering and Computer Science, Vanderbilt University, P. O. Box 351821 Station B, Nashville, TN 37235-1821, USA} \ead{robert.a.weller@vanderbilt.edu}

\begin{keyword}

Geant4 \sep Screened Scattering \sep Coulomb Scattering \sep Elastic Recoil \sep Nuclear Stopping Power \sep NIEL \sep Non-Ionizing Energy Loss \sep TRIM \sep SRIM \sep Displacement Damage

\PACS 34.50 \sep 34.50.B \sep 61.85 \sep 02.50.N \sep 02.70.L  

\end{keyword}

\begin{abstract}

An algorithm has been developed for the \textsc{Geant4} Monte-Carlo package for the efficient computation of screened Coulomb interatomic scattering. It explicitly integrates the classical equations of motion for scattering events, resulting in precise tracking of both the projectile and the recoil target nucleus. The algorithm permits the user to plug in an arbitrary screening function, such as Lens-Jensen screening, which is good for backscattering calculations, or Ziegler-Biersack-Littmark screening, which is good for nuclear straggling and implantation problems. This will allow many of the applications of the TRIM and SRIM codes to be extended into the much more general \textsc{Geant4} framework where nuclear and other effects can be included. 

\end{abstract}

\end{frontmatter}

\section{Introduction}

The necessity of accurately computing the characteristics of interatomic
scattering arises in many disciplines in which energetic ions pass
through materials. Traditionally, solutions to this problem not involving
hadronic interactions have been dominated by the well-known TRIM \cite{TRIM1980}
and SRIM \cite{SRIM1988,SRIMWEB} codes. These codes take a Monte-Carlo
approach to computing distributions of ions passing through a material,
and use a universal parameterized formula to determining a scattering
angle for a particle which collides with a target nucleus. This formula
is a parameterization of the scattering integrals computed from a
parameterization of an interatomic screening function. This approach
is reasonably successful, but not very flexible. In particular, it
is relatively difficult to introduce into such a system a particular
screening function which has been measured for a specific atomic pair,
rather than the universal functions which are applied. 

In recent years, a more general framework, \noun{Geant4} \cite{Geant4NIM2003,GeantWeb},
is being developed by the high-energy physics community for the handling
of the motion of energetic particles through matter. Like TRIM and
SRIM, it takes a Monte-Carlo approach to produce statistical distributions
of particles as they move through various types of matter. However,
the \noun{Geant4} collaboration has developed a much larger toolkit
than SRIM for handling very complex geometries, and for including
many physical processes other than just the traditional electronic
stopping and nuclear stopping in its computation. In many problems
of current interest, such as the behavior of semiconductor device
physics in a space environment, nuclear reactions, particle showers,
and other effects are critically important in modeling the full system.
Thus, it is important to have components in the \noun{Geant4} toolkit
to bridge the gap between the effective handling of low-energy processes
in simple geometries provided by SRIM and the very general framework
for nuclear events and complex geometries already available in \noun{Geant4}.

To be consistent with the general \noun{Geant4} philosophy of providing
tools which are flexible and extensible as better physics models become
available, it was decided that introducing Ziegler-Biersack-Littmark
(ZBL) universal scattering method \cite{ZBLUniversalScattering} into
\noun{Geant4} was inappropriately limiting. Although for most problems
to which SRIM is applied, the accuracy of the universal scattering
equations is sufficient, it is easy to imagine that for situations
in which precise event rates are needed, one might wish to include
high-precision measured interatomic potentials into a computation.
The approach described below makes this very simple, while allowing
one to use the well-established ZBL screening function \cite{ZBLScreeningFn}
(but not the ZBL 'magic formula' for scattering integrals) for the
many situations in which it is sufficiently accurate.

\section{Method}

The method used in this computation is a variant of a subset of the
method described by the authors in a previous paper \cite{MendenhallWellerXSection}.
A very short recap of the basic material is included here. The scattering
of two atoms from each other is assumed to be a completely classical
process, subject to an interatomic potential described by a potential
function\begin{equation}
V(r)=\frac{Z_{1}Z_{2}e^{2}}{r}\phi\left(\frac{r}{a}\right)\label{VR_eqn}\end{equation}
where $Z_{1}$ and $Z_{2}$ are the nuclear proton numbers, $e^{2}$
is the electromagnetic coupling constant ($q_{e}^{2}/4\pi\epsilon_{0}$
in SI units), $r$ is the inter-nuclear separation, $\phi$ is the
screening function describing the effect of electronic screening of
the bare nuclear charges, and $a$ is a characteristic length scale
for this screening. In most cases, $\phi$ is a universal function
used for all ion pairs, and the value of $a$ is an appropriately
adjusted length to give reasonably accurate scattering behavior. In
the method described here, there is no particular need for a universal
function $\phi$, since the method is capable of directly solving
the problem for most physically plausible screening functions. It
is still useful to define a typical screening length $a$ in the calculation
described below, to keep the equations in a form directly comparable
with our previous work even though, in the end, the actual value is
irrelevant as long as the final function $\phi(r)$ is correct. From
this potential $V(r)$ one can then compute the classical scattering
angle from the reduced center-of-mass energy $\varepsilon\equiv E_{c}a/Z_{1}Z_{2}e^{2}$
(where $E_{c}$ is the kinetic energy in the center-of-mass frame)
and reduced impact parameter $\beta\equiv b/a$\begin{equation}
\theta_{c}=\pi-2\beta\int_{x_{{\scriptscriptstyle 0}}}^{\infty}f(z)\, dz/z^{2}\label{theta_eqn}\end{equation}
where\begin{equation}
f(z)=\left(1-\frac{\phi(z)}{z\,\varepsilon}-\frac{\beta^{2}}{z^{2}}\right)^{-1/2}\label{fz_eqn}\end{equation}
and $x_{{\scriptscriptstyle 0}}$ is the reduced classical turning
radius for the given $\varepsilon$ and $\beta$. 

The problem, then, is reduced to the efficient computation of this
scattering integral. In our previous work, a great deal of analytical
effort was included to proceed from the scattering integral to a full
differential cross section calculation, but for application in a Monte-Carlo
code, the scattering integral $\theta_{c}(Z_{1},\, Z_{2},\, E_{c},\, b)$
and an estimated total cross section $\sigma_{{\scriptscriptstyle 0}}(Z_{1},\, Z_{2},\, E_{c})$
are all that is needed. Thus, we can skip algorithmically forward
in the original paper to equations 15-18 and the surrounding discussion
to compute the reduced distance of closest approach $x_{{\scriptscriptstyle 0}}$.
This computation follows that in the previous work exactly, and will
not be reintroduced here.

For the sake of ultimate accuracy in this algorithm, and due to the
relatively low computational cost of so doing, we compute the actual
scattering integral (as described in equations 19-21 of \cite{MendenhallWellerXSection})
using a Lobatto quadrature of order 6, instead of the 4th order method
previously described. This results in the integration accuracy exceeding
that of any available interatomic potentials in the range of energies
above those at which molecular structure effects dominate, and should
allow for future improvements in that area. The integral $\alpha$
then becomes (following the notation of the previous paper)\begin{equation}
\alpha\approx\frac{1+\lambda_{{\scriptscriptstyle 0}}}{30}+\sum_{i=1}^{4}w'_{i}\, f\left(\frac{x_{{\scriptscriptstyle 0}}}{q_{i}}\right)\label{alpha_eq}\end{equation}
where \begin{equation}
\lambda_{{\scriptscriptstyle 0}}=\left(\frac{1}{2}+\frac{\beta^{2}}{2\, x_{{\scriptscriptstyle 0}}^{2}}-\frac{\phi'(x_{{\scriptscriptstyle 0}})}{2\,\varepsilon}\right)^{-1/2}\label{lambda_eqn}\end{equation}
 and \\
$w'_{i}\in${[}0.03472124, 0.1476903, 0.23485003, 0.1860249{]} 

and \\
$q_{i}\in${[}0.9830235, 0.8465224, 0.5323531, 0.18347974{]}\\
 (See appendix \ref{Lobatto}). Then \\
\begin{equation}
\theta_{c}=\pi-\frac{\pi\beta\alpha}{x_{{\scriptscriptstyle 0}}}\label{thetac_alpha_eq}\end{equation}

The other quantity required to implement a scattering process in \noun{Geant4}
is the total scattering cross section $\sigma_{{\scriptscriptstyle 0}}$
for a given incident ion and a material through which the ion is propagating.
This value requires special consideration for a process such as screened
scattering. In the limiting case that the screening function is unity,
which corresponds to Rutherford scattering, the total cross section
is infinite. For various screening functions, the total cross section
may or may not be finite. However, one must ask what the intent of
defining a total cross section is, and determine from that how to
define it. 

In \noun{Geant4}, the total cross section is used to determine a
mean-free-path $l_{\mu}$ which is used in turn to generate random
transport distances between discrete scattering events for a particle.
In reality, where an ion is propagating through, for example, a solid
material, scattering is not a discrete process but is continuous.
However, it is a useful, and highly accurate, simplification to reduce
such scattering to a series of discrete events, by defining some minimum
energy transfer of interest, and setting the mean free path to be
the path over which statistically one such minimal transfer has occurred.
This approach is identical to the approach developed for the original
TRIM code \cite{TRIM1980}. As long as the minimal interesting energy
transfer is set small enough that the cumulative effect of all transfers
smaller than that is negligible, the approximation is valid. As long
as the impact parameter selection is adjusted to be consistent with
the selected value of $l_{\mu}$, the physical result isn't particularly
sensitive to the value chosen. One of the sets of validation tests
discussed below will verify the truth of this hypothesis, and will
determine what reasonable values for this minimal energy transfer
are. 

Noting, then, that the actual physical result isn't very sensitive
to the selection of $l_{\mu},$ one can be relatively free about defining
the cross section $\sigma_{{\scriptscriptstyle 0}}$ from which $l_{\mu}$
is computed. The choice used for this implementation is fairly simple.
Define a physical cutoff energy $E_{min}$ which is the smallest energy
transfer to be included in the calculation. Then, for a given incident
particle with atomic number $Z_{1}$, mass $m_{1}$, and lab energy
$E_{inc}$, and a target atom with atomic number $Z_{2}$ and mass
$m_{2}$, compute the scattering angle $\theta_{c}$ which will transfer
this much energy to the target from the solution of\begin{equation}
E_{min}=E_{inc}\,\frac{4\, m_{1}\, m_{2}}{(m_{1}+m_{2})^{2}}\,\sin^{2}\frac{\theta_{c}}{2}\label{recoil_kinematics}\end{equation}
and then solve, by iterative inversion of eq.~(\ref{thetac_alpha_eq}),
the value of the impact parameter $b$ at which this value of $\theta_{c}$
is achieved. Then, define the total cross section to be $\sigma_{{\scriptscriptstyle 0}}=\pi b^{2}$,
the area of the disk inside of which the passage of an ion will cause
at least the minimum interesting energy transfer. Because this process
is relatively expensive, and the result is needed extremely frequently,
the values of $\sigma_{{\scriptscriptstyle 0}}(E_{inc})$ are precomputed
for each pairing of incident ion and target atom, and the results
cached. However, since the actual result isn't very critical, the
cached results can be stored in a very coarsely sampled table without
degrading the calculation at all, as long as the values of the $l_{\mu}$
used in the impact parameter selection are rigorously consistent with
this table.

The final necessary piece of the scattering integral calculation is
the statistical selection of the impact parameter $b$ to be used
in each scattering event. This selection is done following the original
algorithm from TRIM, where the cumulative probability distribution
for impact parameters is \begin{equation}
P(b)=1-\exp\left(\frac{-\pi\, b^{2}}{\sigma_{{\scriptscriptstyle 0}}}\right)\label{cum_prob}\end{equation}
where $N\,\sigma_{{\scriptscriptstyle 0}}\equiv1/l_{\mu}$ where $N$
is the total number density of scattering centers in the target material
and $l_{\mu}$ is the mean free path computed in the conventional
way. To produce this distribution from a uniform random variate $r$
on (0,1{]}, the necessary function is\begin{equation}
b=\sqrt{\frac{-\log r}{\pi\, N\, l_{\mu}}}\label{sampling_func}\end{equation}
This choice of sampling function does have the one peculiarity that
it can produce values of the impact parameter which are larger than
the impact parameter which results in the cutoff energy transfer,
as discussed above in the section on the total cross section, with
probability $1/e$. When this occurs, the scattering event is not
processed further, since the energy transfer is below threshold. For
this reason, impact parameter selection is carried out very early
in the algorithm, so the effort spent on uninteresting events is minimized.

The above choice of impact sampling is modified when the mean-free-path
is very short. If $\sigma_{{\scriptscriptstyle 0}}>\pi\left(\frac{l}{2}\right)^{2}$
where $l$ is the approximate lattice constant of the material, as
defined by $l=N^{-1/3}$, the sampling is replaced by uniform sampling
on a disk of radius $l/2$, so that\begin{equation}
b=\frac{l}{2}\sqrt{r}\label{flat_sampling}\end{equation}
This takes into account that impact parameters larger than half the
lattice spacing do not occur, since then one is closer to the adjacent
atom. This also derives from TRIM. 

One extra feature is included in our model, to accelerate the production
of relatively rare events such as high-angle scattering. This feature
is a cross-section scaling algorithm, which allows the user access
to an unphysical control of the algorithm which arbitrarily scales
the cross-sections for a selected fraction of interactions. This is
implemented as a two-parameter adjustment to the central algorithm.
The first parameter is a selection frequency $f_{h}$ which sets what
fraction of the interactions will be modified. The second parameter
is the scaling factor for the cross-section. This is implemented by,
for a fraction $f_{h}$ of interactions, scaling the impact parameter
by $b'=b/\sqrt{scale}$. This feature, if used with care so that it
does not provide excess multiple-scattering, can provide between 10
and 100-fold improvements to event rates. If used without checking
the validity by comparing to un-adjusted scattering computations,
it can also provide utter nonsense. This scaling provides a mechanism
to address the issues discussed in the recent literature \cite{ERDMonteCarlo2004,FastMonteCarlo2004}.
In particular, it solves the problem of preserving multiple scattering
effects by leaving most interactions unaffected (if the fraction $f_{h}$
is much less than unity) while still providing significantly enhanced
yields of hard collisions.

\section{Validation}

There are a number of features of this model which need to be verified
against good physical data and theoretical expectations, within the
\noun{Geant4} framework, to assure the compatibility of its statistical
sampling methods with those of \noun{Geant4}, the correctness of
the underlying physical assumptions, and the freedom of the implemented
code from logic defects. 

In many of the comparisons below, we use data from SRIM as the reference.
This choice was made because of the wide acceptance of SRIM in many
fields as a useful and well-tested tool. Furthermore, the large compendium
of measurements referenced at \cite{SRIMWEB} provides a central repository
for such information. However, precise agreement with SRIM is not
the goal of these comparisons. The method we describe should be more
accurate than the nuclear scattering component of SRIM, in that it
exactly integrates the scattering, rather than using a universal,
parameterized approximation to it. Further, this method can be applied
with specific internuclear potentials, allowing further enhancement
in accuracy over SRIM. However, since electronic stopping contributes
to all the processes discussed below, and SRIM is highly optimized
for this, especially in compound materials, it will most likely provide
different, and probably better, results in some domains.

In the results below, the smooth curves listed as SRIM data are digitized
from the curves on the graphs provided on the SRIM website as of May,
2004. The digitized data was oversampled, and then smoothed via least-squares
cubic splining. If a reader of this work intends to digitize data
from our graphs, it is highly recommended that the SRIM data be obtained
from the original, and not re-digitized from our copy, to prevent
accumulation of errors.

\subsection*{Rutherford Scattering}

\begin{center}\begin{figure}[ht]\caption{This compares the backscattering of 2 MeV $\alpha$ particles from a 100 nm Si foil to the theoretical value.  A total of $10^8$ particles were used, with a cross-section enhancement of 100, to get these statistics.  The data are binned in $\cos\theta$ bins with $\Delta\cos\theta=0.02$. The points are plotted at the bin center, transformed to an angle. Computing time on a modern laptop computer is a few hours. }\label{RBSfig}\includegraphics[%
  clip,
  width=1.0\columnwidth,
  keepaspectratio]{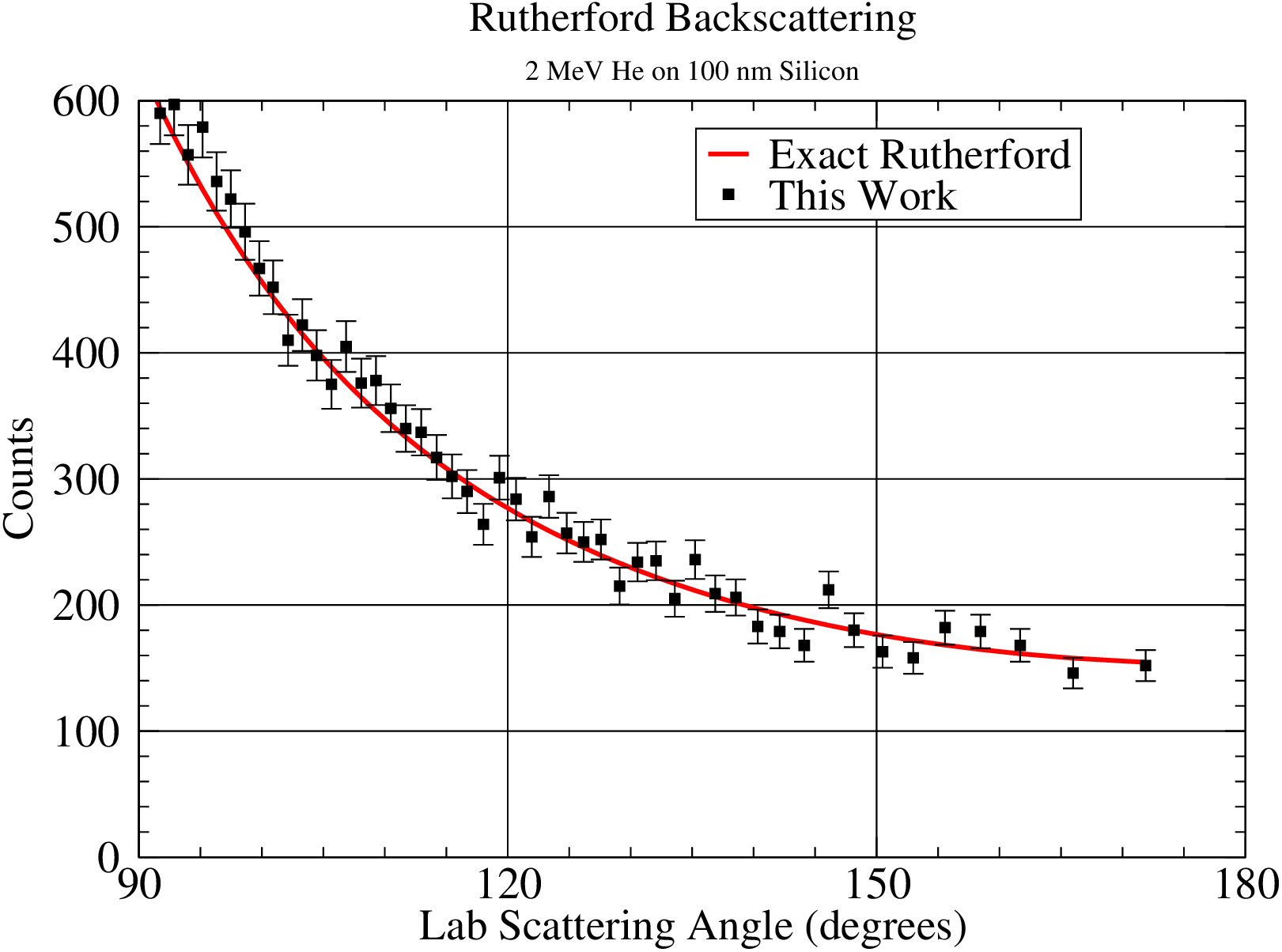}\end{figure}\end{center}

\begin{center}\begin{figure}[ht]\caption{Alpha particle range in air.  Even at the lowest energies for which data are available, this process is dominated by electronic stopping.  Thus, the differences which errors in our calculation would introduce here are quite small.}\label{He_air_fig}\includegraphics[%
  width=1.0\columnwidth]{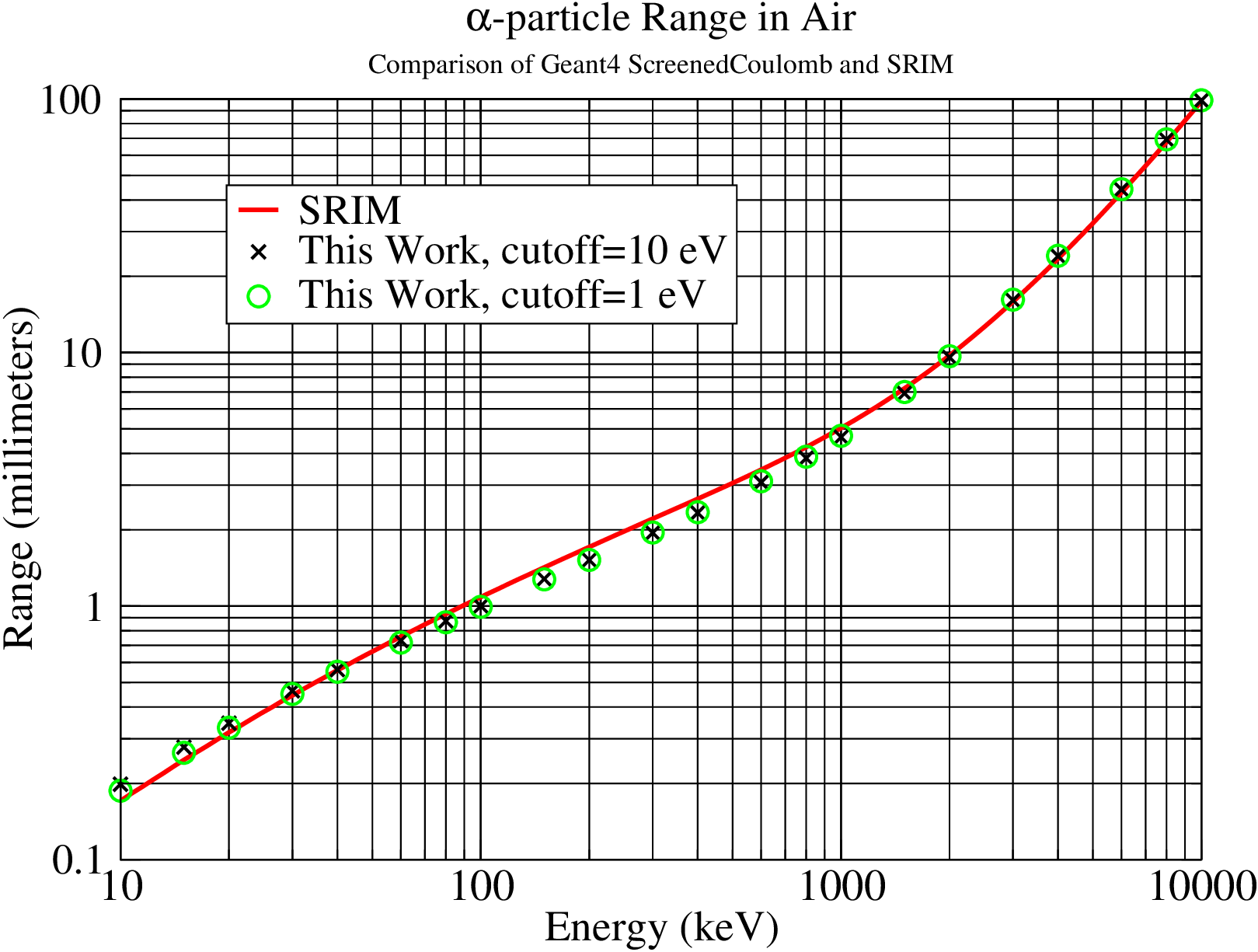}\end{figure}\end{center}

The first, and most basic, test is to see if the code replicates well-known
scattering distributions. It is important to note that, for this method,
the Rutherford cross-section is not a special case, insofar as the
integrand in eq.(\ref{theta_eqn}) does not reduce to an unusually
simple form under our selected change-of-variable. Thus, agreement
with the expectations of Rutherford scattering, under conditions where
this cross-section is accurate, is a strong test of validity. To make
this comparison, a very typical set of scattering parameters was chosen,
which is the scattering of 2 MeV $\alpha$ particles from a 100~nm
thick silicon foil. The results shown in Figure \ref{RBSfig} were
computed both for the unmodified cross-section, and for a cross-section
increased by a factor of 100, as described above. The excellent agreement
between the two results indicates that the scaling process is not
distorting the calculation. The scaled cross section is used in the
final statistical comparison, since the total number of events is
much larger, resulting in better statistics. In the case of this calculation,
where the scattering probability from such a thin target is very small
for almost all scattering angles greater than 10~degrees, and multiple
scattering provides no significant contribution, this agreement, even
with a large cross-section scaling, is expected. Note that the calculation
has been cut off for small scattering angles, where deviations from
the Rutherford cross-section arise and where this assumption about
multiple scattering is not valid. Other validation tests will cover
this region.

\subsection*{Ion Implantation and Range Straggling}

\begin{center}\begin{figure}[ht]\caption{Boron range and straggling in silicon. At high energies, where most of the energy loss is electronic, straggling is small and the implantation profile is very narrow.}\label{Boron_si_fig}\includegraphics[%
  width=1.0\columnwidth]{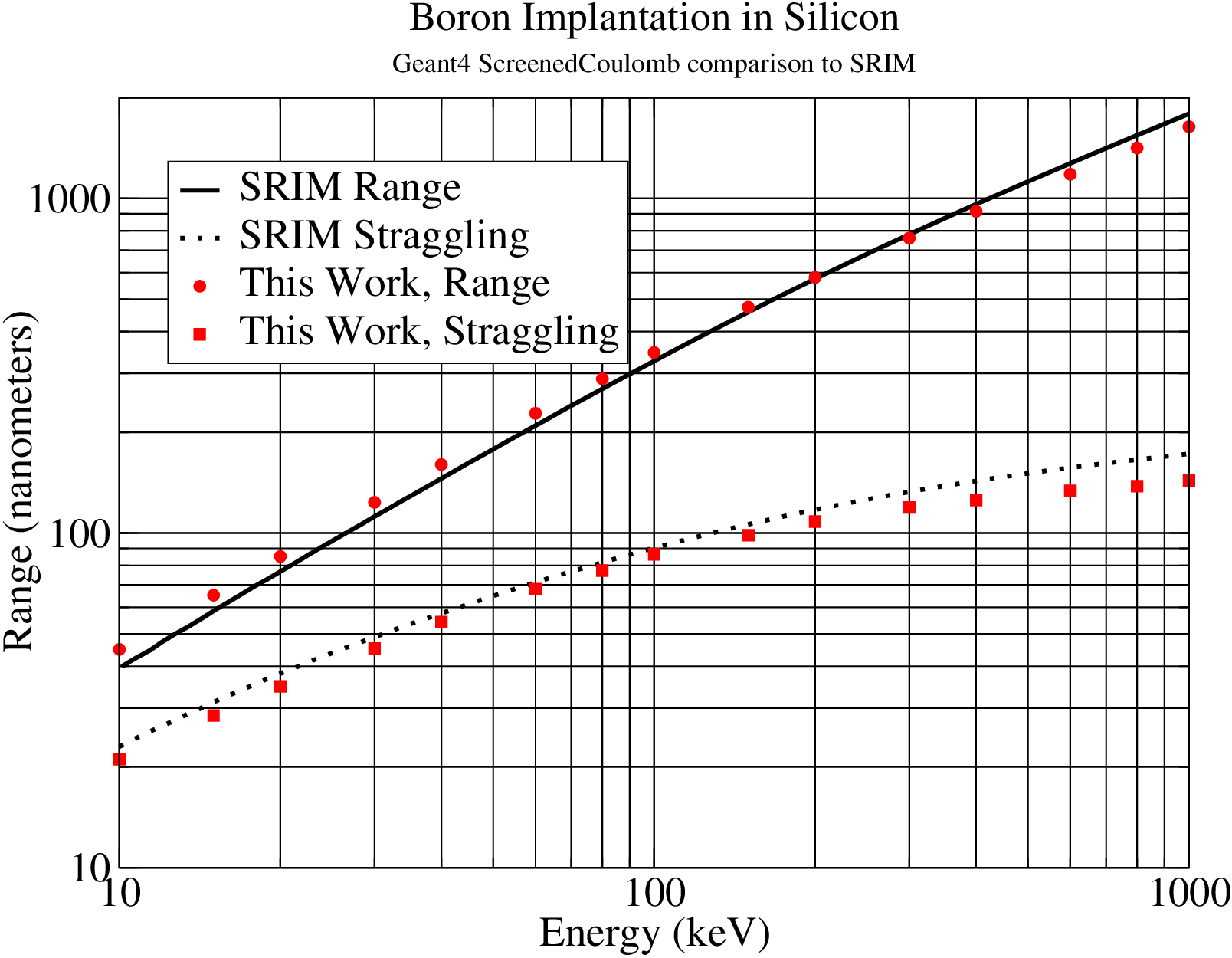}\end{figure}\end{center}

\begin{center}\begin{figure}[ht]\caption{Arsenic range and straggling in silicon.  At the relatively low velocities associated with the kinetic energy range covered by this data set, nuclear scattering is strong, so the straggling is a large fraction of the projected range and the process described in this work is extremely important.}\label{Arsenic_si_fig}\includegraphics[%
  width=1.0\columnwidth]{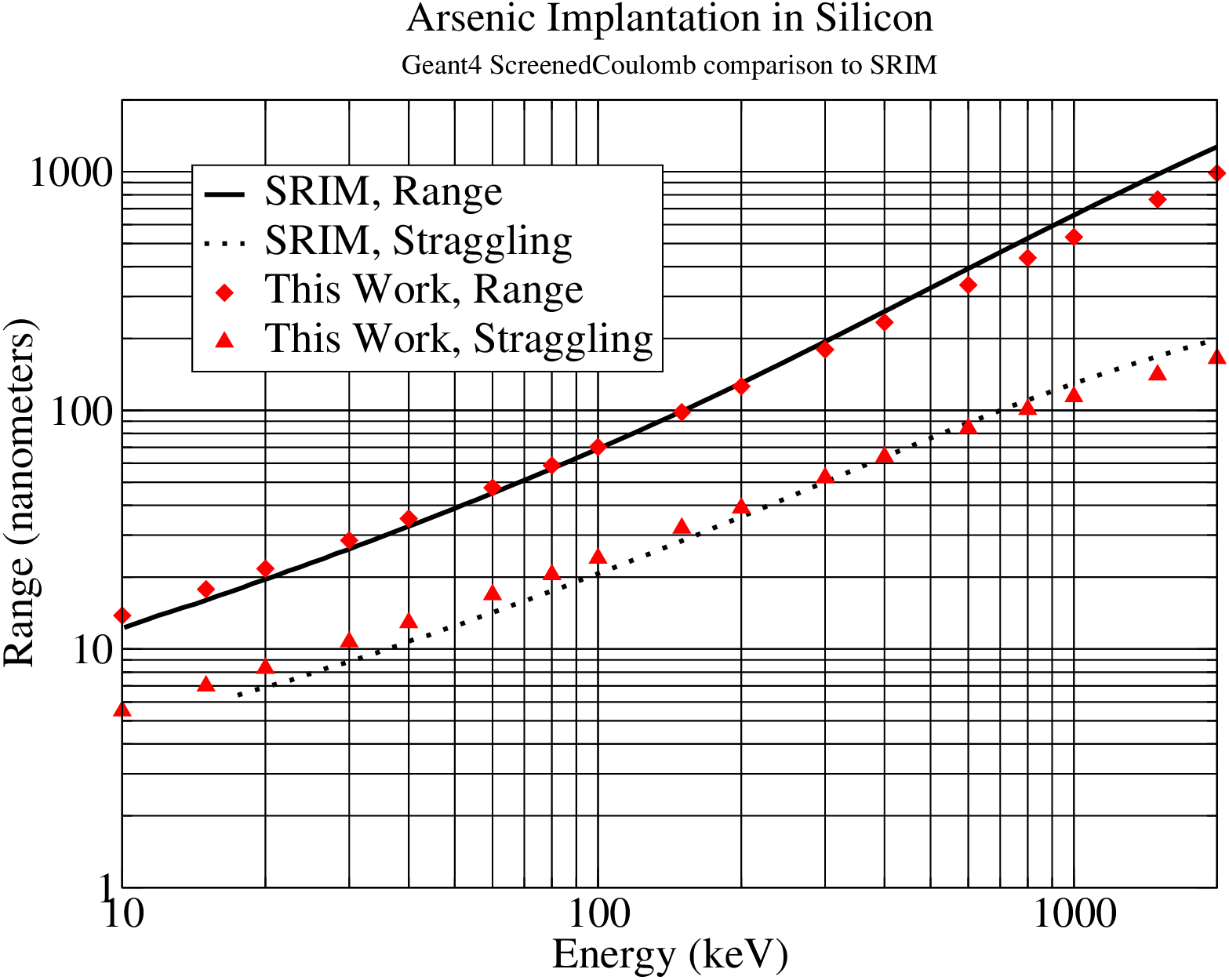}\end{figure}\end{center}

To test the process described here for accuracy in final stopping
situations, we will compare with two very well documented systems:
the propagation of $\alpha$ particles in air, shown in figure \ref{He_air_fig},
which has been studied since 1913 (see, \emph{e.g.,} \cite{Taylor1913}
and an extensive bibliography at \cite{SRIMWEB}), and the implantation
of dopants in silicon, which has been thoroughly studied because of
its importance to the semiconductor industry. The two cases for implantation
also cover two very different regions of parameter space. The implantation
of boron in silicon involves a light projectile on a heavy target,
where some high-angle scattering events can occur. Conversely, for
arsenic in silicon, where the projectile is much heavier than the
target, all scattering is very forward. Also, the arsenic-in-silicon
test case samples data down to extremely low velocity. These are shown
in figures \ref{Boron_si_fig} and \ref{Arsenic_si_fig}.

\subsection*{Forward Multiple Scattering}

The ability of this process to correctly produce small-angle forward
scattering is likely to be quite important. Typically, one is interested
in the diffusion of an incoming beam particle as it is transported
through, for example, a gas cell or metal vacuum window. Although
\noun{Geant4} includes its own multiple scattering process to approximate
this, that process is optimized for efficiency with very high energy
ions, and for cases in which such scattering is quite weak. 

\begin{center}\begin{figure}[ht]\caption{Forward scattering of 270 keV $\alpha$ particles from a $100\; \mu\textrm{g}/\textrm{cm}^2$ carbon foil.}\label{MSFfig}\includegraphics[%
  clip,
  width=1.0\columnwidth,
  keepaspectratio]{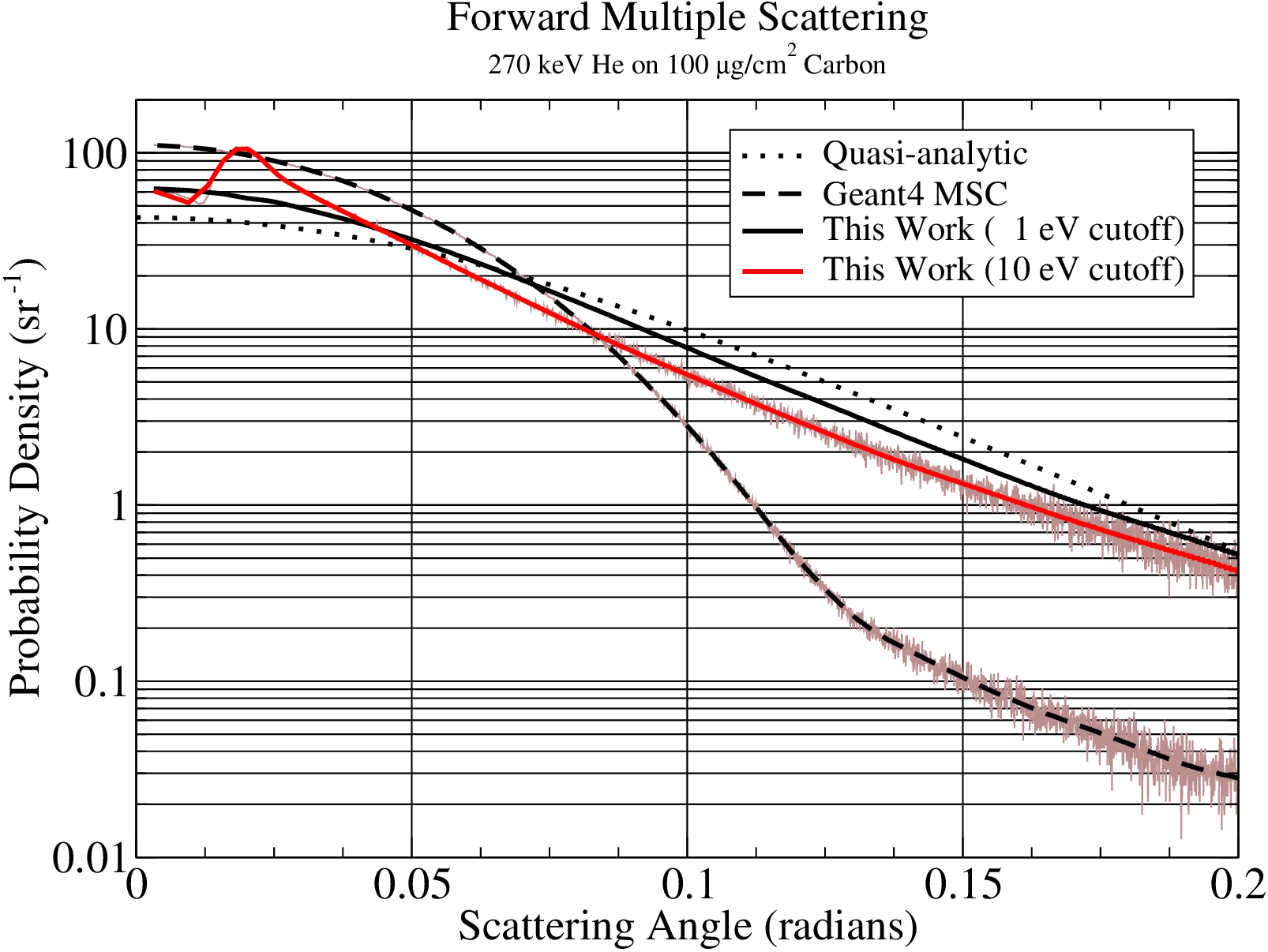}\end{figure}\end{center}

Since the authors of this work have previous experience with and interest
in scattering of 270 keV $\alpha$ particles for surface analysis,
we chose a test case from this domain. The data presented in Figure
\ref{MSFfig} show the results. On two of the curves, representative
raw data from the Monte-Carlo simulation are shown, to illustrate
typical counting statistics for the data from which these curves are
computed. The actual smooth curves are least-squares cubic splines
of the data.

The effect of the low-energy cutoff of this process is apparent at
very small angles, where scattering is suppressed below about 0.02
radians. This should be considered as a warning to users of this process,
in that one must consider carefully what scattering angles are of
interest. Choosing the cutoff energy too high will result in distortion
of the scattering distribution for small angles, but can result in
improved computational speed. In the case of this foil, the mean-free-path
is about 30\% of the foil thickness for the 10~eV cutoff, and about
5\% of the foil thickness for the 1~eV cutoff. It is likely that
this is a good way to estimate an appropriate cutoff energy, since
good multiple scattering accuracy depends upon at least a few interactions
being applied to each ion passing through the target. With the cutoff
at 10~eV, for this foil thickness, one is primarily sampling ions
which have not undergone many scatterings in the foil, so the distribution
is somewhat too narrow. By the time the cutoff has been reduced to
1~eV, each ion is scattered about 20 times in the foil, and the resulting
statistics can be expected to be quite accurate.

The curve labeled 'Quasi-analytic' is computed using the method of
\cite{MendenhallWellerMSC}, adapted for the case where the energy
of the beam exiting the target is very different from the initial
energy. This technique is a small-angle approximation to the multiple
scattering, but it uses high-accuracy scattering cross-sections computed
directly from the screening functions. 

The curve labeled 'Geant4 MSC' uses the G4MultipleScattering (MSC)
process which is a standard part of the \noun{Geant4} package. This
process uses a statistical approach to multiple scattering \cite{Lewis1950MSC},
somewhat similar to that of \cite{MendenhallWellerMSC}, but which
is based on universal, parameterized scattering cross sections. As
can be seen from the graph, it severely underestimates high-angle
scattering for heavy ions. More significant, and one of the primary
reasons for this work, is that it does not produce recoil particles.

\begin{figure}[ht]\caption{Forward scattering of 270 keV protons from a $100\; \mu\textrm{g}/\textrm{cm}^2$ carbon foil. In this case, agreement between MSC and this work is  close.}\label{msc_protons}\includegraphics[%
  width=1.0\columnwidth,
  keepaspectratio]{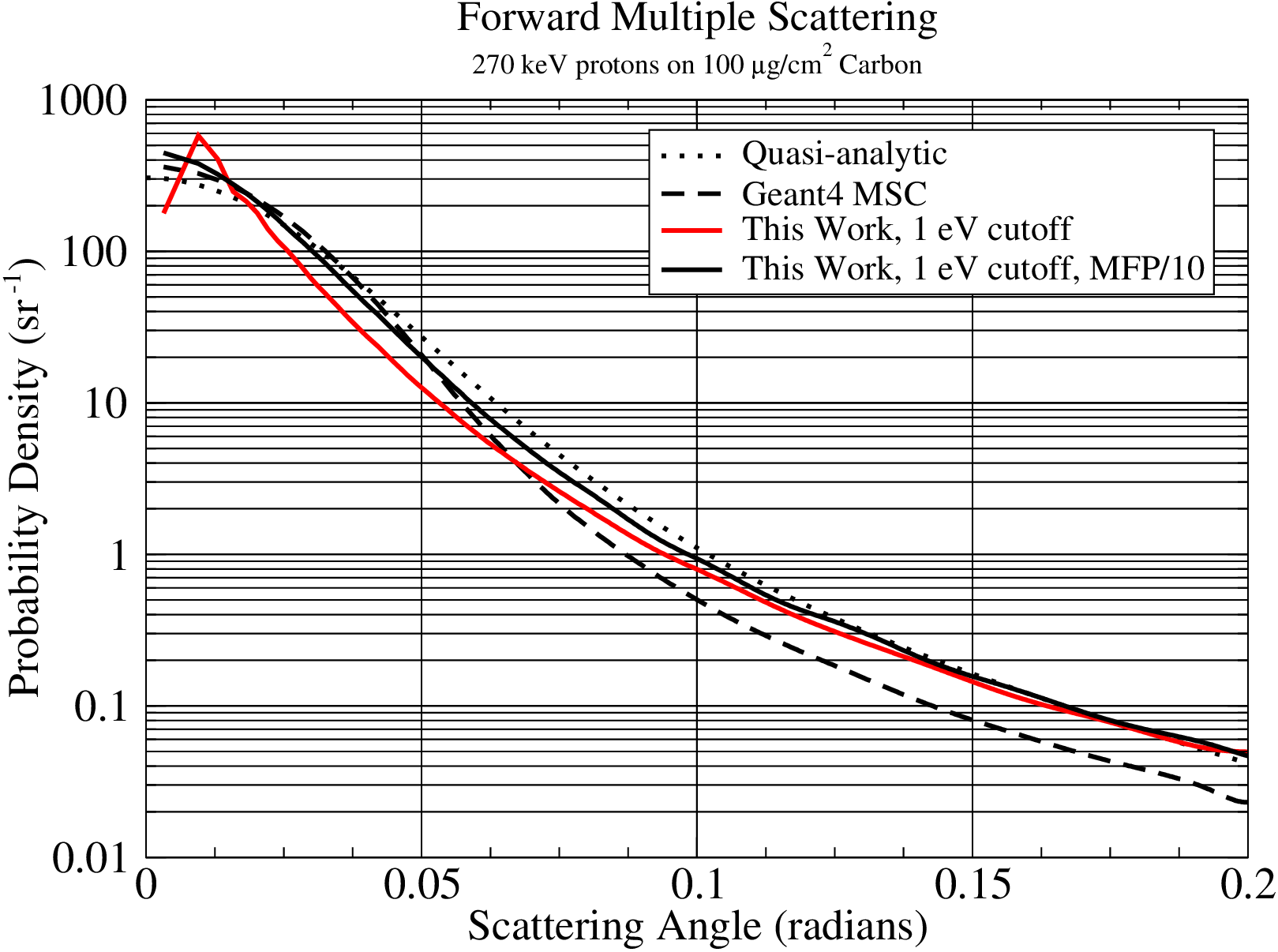}\end{figure}

Figure \ref{msc_protons} shows another forward multiple-scattering
case, with two important differences from that in Figure \ref{MSFfig}.
First, the intrinsic \noun{Geant4} multiple scattering works quite
well for computing the width of the forward scattering distribution
for protons. Note that, even for this case, it shows a deficit of
almost a factor of two for higher-angle scattering. Second, another
parameter of the the screened Coulomb model is exercised in this run. 

In the discussion above, the mean free path was adjusted to be substantially
less than the target thickness by adjusting the lower cutoff energy
of the data tables being built. In the case of protons at this energy,
even with a 1~eV cutoff energy, the mean free path is too long to
get good results. Lowering the physics cutoff below 1~eV is probably
unphysical, since the potentials aren't well known far into the tails,
and solid state effects probably make them meaningless. However, the
real problem with the long mean free path is not that any physics
is missing; it is that many particles get through with zero scattering
events, since the mean number of events is only a few. 

The model contains a parameter MFPScale which allows the user to artificially
shorten the mean free path in a consistent manner, so that more attempts
are made to scatter and the Poisson statistics don't overlap zero
events very strongly. This allows the user to adjust the model for
the case of very thin foils. The curve labeled 'MFP/10' has had the
mean free path reduced a factor of 10, and this fully resolves (at
the expense of increased computing time) the shape of the scattering
distribution near the origin.

\section{Conclusions}

The algorithm presented here provides an accurate and efficient way
to include the effects of Coulomb scattering and the generation of
recoil particles in \noun{Geant4} simulations. This algorithm has
the flexibility to permit the user to select screening functions appropriate
for specific applications, and to provide new screening functions,
without any rewriting of the core code. It should find wide applicability
in many problems in the interaction of fast ions with materials, where
the average multiple-scattering already provided by \noun{Geant4}
must be replaced by a detailed model.

Some improvements may be possible in the future, with appropriate
checks for physical validity. It is the belief of the authors that
the weighting of the target atom selection could be improved with
extra material information. In a material in which atoms live in very
different environments, weighting the selection by the mean-square
bond length for each specific species may improve the statistical
selection accuracy. This would provide some compensation for an atom
being closely caged by its neighbors, such that it is hard to hit
that atom without also making a close pass to other species. Such
weighting would have to be implemented by providing extra information
for the target material, beyond that which \noun{Geant4} normally
uses. A mechanism for this already exists in \noun{Geant4}. 

Although the current implementation of the algorithm is strictly non-relativistic,
the authors are investigating simple extensions which will correctly
handle small-angle collisions for relativistic incident particles
and non-relativistic recoils. These collisions are important for nuclear
stopping power and Non-Ionizing Energy Loss (NIEL) calculations at
high energies. Hard collisions at relativistic energies are not to
be included, since such collisions have a small Coulombic cross-section
and will be combined hadronic-Coulombic events, in which case the
screening function and classical scattering concepts are not valid,
and the resulting processes are handled by nuclear-reaction codes. 

\begin{description}
\item [Acknowledgment:]The authors would like to thank the members of the
\noun{Geant4} core physics team, especially Maria Grazia Pia and
Hans-Peter Wellisch, for helpful discussions.
\end{description}
\appendix

\section{Appendix: Derivation of Integration Constants}

\label{Lobatto}The coefficients for the summation to approximate
the integral for $\alpha$ in eq.(\ref{alpha_eq}) are derived from
the values in Abramowitz \& Stegun \cite[sec. 25.4.32 and table 25.6]{AbramowitzStegunLobatto},
altered to make the change-of-variable used for this integral. There
are two basic steps to the transformation. First, since the provided
abscissas $x_{i}$ and weights $w_{i}$ are for integration on {[}-1,1{]},
with only one half of the values provided, and in this work the integration
is being carried out on {[}0,1{]}, the abscissas are transformed as:\begin{equation}
y_{i}\in\left\{ \frac{1\mp x_{i}}{2}\right\} \label{as_yxform}\end{equation}
 Then, the primary change-of-variable is applied resulting in:\begin{eqnarray}
q_{i} & = & \cos\frac{\pi\, y_{i}}{2}\label{q_xform}\\
w'_{i} & = & \frac{w_{i}}{2}\sin\frac{\pi\, y_{i}}{2}\label{wprime_xform}\end{eqnarray}
 except for the first coefficient $w'_{1}$where the $\sin()$ part
of the weight is taken into the limit of $\lambda_{{\scriptscriptstyle 0}}$
as described in eq.(\ref{lambda_eqn}). This value is just $w'_{1}=w_{1}/2$.

\section{Appendix: \noun{Geant4} Implementation}

The process being described in this work is a process of class G4DiscreteProcess
in the \noun{Geant4} class hierarchy. As such, it really only needs
to provide a few functions to interact correctly with the \noun{Geant4}
world. The functions used by \noun{Geant4} are the constructor,
GetMeanFreePath(), and PostStepDoIt(). Internally, the functions are
divided into two classes, a private CrossSection class which handles
loading of screening tables and total cross-section tables, and the
main Screened\-Nuclear\-Recoil class, which implements the required
G4Discrete\-Process interface. There are a few other utilities provided
which allow such functions as adjusting various model cutoffs, cross-section
biasing, and control of energy deposition. The computation and caching
of screening functions and total cross-sections is left to an external
\noun{Python} \cite{PythonWeb} program.

\subsection*{Class CrossSection}

This is a class derived from G4\-Cross\-Section\-Handler, and provides
extensions to that class to read data in via a pipe from an external
process, and to store screening tables along with cross section tables.

\subsubsection*{Method: LoadData( G4String screeningKey, G4int z1, G4double m1, G4double
recoilCutoff)}

This method is the primary reason for the existence of this class.
The standard G4\-Cross\-Section\-Handler class is designed to read
precomputed cross sections from stored text files in a fixed format.
For the purposes of this system, it is not only necessary to have
cross sections, but screening tables, and these tables depend on the
value of the minimum scattering energy cutoff, as described above.
The number of available parameters would result in a combinatorially
large number of files being required in the database. To avoid this
problem, the author decided that it was more efficient to use a small,
external program, written in the \noun{Python} programming language,
to dynamically generate the files as needed and to maintain a cache
of the tables actually used for quick re-use. This permits the user
to include custom screening functions by adding them to a small, easily
maintained \noun{Python} module. This module returns data to the
main program through a \noun{Unix}-style pipe interface, which is
supported on all POSIX-compatible platforms, and appears as a file
to both the calling and called program.

\subsubsection*{Method: SelectRandomTargetUnweighted()}

This method selects an atom from the currently active material, based
only on the stoichiometry of the material. Weighting the selection
by the scattering cross section would result in double-counting the
weight, so it is not done.

\subsection*{Class ScreenedNuclearRecoil}

The methods documented below are the public methods of the class which
are directly useful to the end user for setting physics parameters.
The main PostStepDoIt() method implements the algorithm described
in the rest of the paper, and is only used by \noun{Geant4} internals.

\subsubsection*{Method: ScreenedNuclearRecoil( const G4String\& processName = \char`\"{}Screened\-Elastic\char`\"{},
const G4String \&ScreeningKey = \char`\"{}zbl\char`\"{}, G4bool GenerateRecoils
= 1, G4double RecoilCutoff = 100.0{*}eV, G4double PhysicsCutoff =
10.0{*}eV)}

The constructor for this process allows the user to set a number of
important physics parameters. 

\begin{description}
\item [ScreeningKey]selects which screening function will be requested
from the external \noun{Python} module which generates screening
tables and cross sections
\item [GenerateRecoils]controls whether recoil particles are generated
and tracked, or whether a local energy deposition is made with the
energy that would otherwise have been transferred to a recoil particle.
\item [RecoilCutoff]sets the energy below which a recoil will not be generated,
and below which an incoming particle will be stopped with no further
interaction. The stopped particles deposit energy (if permitted),
and are allowed to decay if appropriate ({}``stop-but-alive'' if
any atRest processes exist, otherwise {}``stop-and-kill'').
\item [PhysicsCutoff]sets the energy cutoff used in the calculation of
the total scattering cross-section, as described above. Its value
is typically set to between 1~eV and 10~eV for problems in which
forward multiple scattering is important. For problems involving backscattering,
it can be raised to 100~eV or beyond to improve efficiency. Changing
this parameter changes the mean-free-path and should not have a strong
effect, unless the mean-free-path is approaching the length scale
of the material in which the particle is traveling.
\end{description}

\subsubsection*{Method: AllowEnergyDeposition( G4bool flag)}

If this is called with a flag of zero or false, all calls to deposit
local energy are suppressed, but all processes proceed normally otherwise.
This is useful for measuring how much energy is deposited as a result
of nuclear collisions. Note that this leaves the rest of the physics
and random number consumption strictly alone, so that by resetting
the random number generation, one can run exactly the same events
with this on and off, and subtract the results to see how much energy
was deposited in final stopping of particles when they reach the \noun{RecoilCutoff}
energy described above.

\subsubsection*{Method: EnableRecoils( G4bool flag)}

This dynamically controls the same variable set by \noun{GenerateRecoils}
in the constructor. The value can be changed at any time.

\subsubsection*{Method: SetMFPScaling( G4double scale)}

This allows the mean-free-path computed from $l_{\mu}=1/N\sigma_{{\scriptscriptstyle 0}}$
to be scaled by an arbitrary amount, in a consistent way so that the
underlying physics isn't changed. It is intended to improve tracking
of particles in thin foils, where the thickness of the foil is less
than a few times the mean-free-path. It can be changed at any time
(e.g. one could add to a stepping action code to change it in a thin
foil, and then reset it for better efficiency in regions with longer
scale lengths).

\subsubsection*{Method: SetRecoilCutoff( G4double energy) }

This dynamically controls the same variable set by \noun{RecoilCutoff}
in the constructor, and can be changed at any time.

\subsubsection*{Method: SetPhysicsCutoff(G4double energy)}

This dynamically controls the same variable set by \noun{PhysicsCutoff}
in the constructor. Although it can be changed at any time, there
is a relatively high cost associated with doing so, since the physics
tables for this process must then be reloaded. It is intended to allow
the user to change it between runs, without restarting the \noun{Geant4}
kernel.

\subsubsection*{Method: SetCrossSectionHardening(G4double fraction, G4double HardeningFactor)}

This enables the cross-section enhancement algorithm described above.
A subset of the interactions, with probability \noun{fraction},
has its cross-section increased by a scale of \noun{HardeningFactor}
by reducing the impact parameter appropriately. It can be changed
at any time.

\subsubsection*{Method: G4double GetNIEL()}

This returns the total energy which has been deposited as local energy
depositions by this process in the most recent step. It is reset at
the start of each step, so it must be accessed in the User\-Stepping\-Action
or in a subclass which overriddes PostStepDoIt() if its value is of
interest. It is assumed to represent the Non-Ionizing Energy Loss
(NIEL) if the \noun{RecoilCutoff} is set low enough that essentially
all of the remaining energy in a particle will be deposited collisionally.
This value is valid whether \noun{AllowEnergyDeposition} is true
or false.

\bibliographystyle{elsart-num}
\bibliography{refs}

\end{document}